\documentclass[aps,twocolumn,pra,
preprintnumbers,amsmath,amssymb,superscriptaddress]{revtex4}
\usepackage{graphicx}
\usepackage{color}

\definecolor{refkey}{rgb}{0.9451,0.2706,0.4941}
\definecolor{labelkey}{rgb}{0.9451,0.2706,0.4941}
 
\pdfoutput=1

\newcommand{\bea}{\begin{align}}
\newcommand{\eea}{\end{align}}

\newcommand{\tr}{\mathrm{Tr}}

\newcommand{\eq} {equation}
\newcommand{\eqa} {eqnarray}
\newcommand{\NN} {\nonumber}
\newcommand{\keff}{k_{\rm eff}}

\begin{document}

\preprint{OU-HET-875}
\preprint{RIKEN-STAMP-19}
\preprint{WIS/07/15-OCT-DPPA}
\preprint{YITP-15-90}

\title{Exact Path Integral for 3D Higher Spin Gravity}

\date{\today}

\author{Masazumi Honda}\email[]{masazumi.honda@weizmann.ac.il} 
\affiliation{{\it Department of Particle Physics, Weizmann Institute of Science, Rehovot 7610001, Israel}}

\author{Norihiro Iizuka}\email[]{iizuka@phys.sci.osaka-u.ac.jp} 
\affiliation{{\it Department of Physics, Osaka University, Toyonaka, Osaka 560-0043, JAPAN}}

\author{Akinori Tanaka}
\email[]{akinori.tanaka@riken.jp}
\affiliation{{\it Interdisciplinary Theoretical Science Research Group, 
RIKEN, Wako 351-0198, JAPAN}}

\author{Seiji Terashima}
\email[]{terasima@yukawa.kyoto-u.ac.jp}
\affiliation{{\it Yukawa Institute for Theoretical Physics, 
Kyoto University, Kyoto 606-8502, JAPAN}}

\begin{abstract}
Extending the works arXiv:1504.05991 and arXiv:1510.02142, we study three dimensional Euclidean higher spin gravity with negative cosmological constant. This theory can be formulated in terms of $SL(N,\mathbb{C})$ Chern-Simons theory. By introducing auxiliary fields, we rewrite it in a supersymmetric way and compute its partition function exactly by using the localization method. We obtain a good expression for the partition function in terms of characters for the vacuum and primaries in 2D unitary CFT with $\mathcal{W}_N$ symmetry. We also check that the coefficients of the character expansion are positive integers and exhibit Cardy formula in the large central charge limit.
\end{abstract}

\maketitle



\noindent

\section{Introduction}

3D quantum gravity is always a fascinating toy model of the realistic 4D quantum gravity.
In \cite{Iizuka:2015jma, Honda:2015hfa}, 
the authors considered 3D pure AdS gravity partition function. 
It is well known that
3D Euclidean Einstein gravity with negative cosmological constant can be formulated 
in terms of the Chern-Simons (CS) theory 
with the gauge group $SL(2,\mathbb{C})$ \cite{Witten:1988hc,Achucarro:1987vz}. 
Furthermore, by introducing only auxiliary fields, this theory can be written in a supersymmetric way.  
Based on this observation, the authors applied the localization technique 
to the 3D Euclidean Einstein gravity, 
and under certain assumptions, obtained the quantum gravity partition function 
\cite{Iizuka:2015jma} and arrived at plausible CFT interpretations \cite{Honda:2015hfa}.

Similarly, 
the CS theory with the gauge group $SL(N,\mathbb{C})$
describes 3D higher spin gravity with negative cosmological constant,
where gravity is coupled to symmetric tensors of spin $3,4,\cdots ,N$ \cite{Blencowe:1988gj}.  
Then it is very natural to generalize our previous works \cite{Iizuka:2015jma} and \cite{Honda:2015hfa}, where 
the gauge group is $SL(2,\mathbb{C})$, to $SL(N,\mathbb{C})$ 
for exact path integral of the higher spin gravity. 
In fact, in our previous work \cite{Honda:2015hfa} 
for the $SL(2,\mathbb{C})$ case, 
we have seen that in the semi-classical limit,  
we can express the exact partition function as summation over 
vacuum and primary characters. 
In this paper, we will see that this nice expression holds even in the higher spin gravity case 
where the gauge group for the CS theory is $SL(N,\mathbb{C})$ and the partition function is written in terms of characters of 2D $\mathcal{W}_N$ CFT.

Another motivation to study the 3D higher spin gravity is 
the recent progress on the Vasiliev higher spin theory \cite{Vasiliev:2003ev}, 
which has brought much attention in the context of AdS/CFT correspondence.
In contrast to usual AdS/CFT, 
the Vasiliev theory is expected to be dual to vector-like CFT as discussed earlier in \cite{Klebanov:2002ja}.
While string theory at high energy has been suspected to have higher spin symmetry \cite{Gross:1988ue},
spectrum of the Vasiliev theory 
is fairly different from the one of typical string theory
and how 
the Vasiliev theory is related to string theory 
is still mysterious.
Furthermore it is known that 
in a black hole solution of the Vasiliev theory,
horizon itself is a gauge dependent concept (see e.g. \cite{Ammon:2012wc}).
Thus the higher spin gravity has various unusual properties and
should be studied more quantitatively.

The Euclidean action of the 3D higher spin gravity is given by
\begin{\eq}
S_{\rm HigherSpin}
=\frac{ik}{4\pi} S_{\rm CS} [\mathcal{A}] -\frac{ik}{4\pi} S_{\rm CS} [\bar{\mathcal{A}}] \,,
\label{eq:action}
\end{\eq}
where
\begin{\eq}
S_{\rm CS} [\mathcal{A}]
= \int_{\mathcal{M}} \tr \left( \mathcal{A}d\mathcal{A} +\frac{2}{3}\mathcal{A}^3  \right) \,.
\end{\eq}
In this paper, we consider the case, where
the subgroup $SL(2,\mathbb{C})\subset SL(N,\mathbb{C})$
is embedded in the principal embedding \footnote{
See e.g. appendix A in \cite{Castro:2011iw} for the definition of the principal embedding.
}. 
Then the gauge field is related to the generalized vielbein $e$ and spin connection $\omega$ by
\begin{\eq}
\label{eq:eomegaArelation}
e=\frac{\ell}{2 i}(\mathcal{A}-\bar{\mathcal{A}}) \,, \quad \omega =\frac{1}{2}(\mathcal{A}+\bar{\mathcal{A}}) \,.
\end{\eq}
where $\ell$ is AdS scale. 
One can denote the metric and higher spin fields 
in terms of trace invariants of the vielbein \cite{Campoleoni:2010zq,Campoleoni:2011hg} such as $g_{\mu\nu} \sim \tr(e_\mu e_\nu)$, 
$\phi_{\mu\nu\lambda} \sim \tr(e_{(\mu} e_\nu e_{\lambda)})$,  
for example. 
The CS level $k$ is related to the Newton constant by $k= 3\ell /2G_N N(N^2 -1)$. 
By generalizing the Brown-Henneaux's work \cite{Brown:1986nw},
we can show that 
central charge $c_Q$ of asymptotic symmetry group in the higher spin gravity 
is given by \cite{Henneaux:2010xg,Campoleoni:2010zq}
$c_Q =k N (N^2 -1) =3\ell /2G_N$.   
The quantization condition for the $SL(N,\mathbb{C})$ group is $k\in \mathbb{Z}$ 
(see e.g.~\cite{Witten:1989ip}). 
However the isometry of $AdS_3$ is $SO(3,1)$, which is locally $SL(2,\mathbb{C})\subset SL(N,\mathbb{C})$. 
If we considered the CS theory with the gauge group $SO(3,1)$, 
then the quantization condition is $k/4 \in\mathbb{Z}$ \cite{Witten:2007kt}. 
Since $SL(N,\mathbb{C})$ group corresponds to the theory of Einstein gravity ($s=2$) with higher spin fields ($2 < s \le N$) and 
we can always set the higher spin fields to be zero, 
we expect that the correct quantization for $k$ is still $k/4 \in\mathbb{Z}$ even for $N > 2$, and 
this leads to $c_Q /24\in\mathbb{Z}$. 
We will see later that the condition $c_Q /24\in\mathbb{Z}$ guarantees that 
convergent and modular invariant partition function does not vanish.

In this paper we exactly compute the partition function of the higher spin gravity
\begin{\eq}
\label{eq:higherspinpartitionZ}
Z_{\rm HigherSpin} 
=\int \mathcal{D}e \mathcal{D}\omega\ e^{-S_{\rm HigherSpin}} ,
\end{\eq}
by generalizing the previous work for the pure Einstein gravity corresponding to $SL(2,\mathbb{C})$ \cite{Iizuka:2015jma}.
For this purpose,
we shall first define the path integral measure for the higher spin gravity in terms of
the Chern-Simons formulation.
In the path integral of the higher spin gravity,
$e$ and $\omega$ can take 
all possible values, corresponding to all  
possible bulk topologies,  
with a given asymptotic AdS boundary condition. Through the relation \eqref{eq:eomegaArelation}, this data for $e$ and $\omega$ is mapped into the data for the gauge bosons $\mathcal{A}$ and $\bar{\mathcal{A}}$. 
On the other hand, in the usual quantum field theory without gravity,  
we shall regard two Chern-Simons theories living on the different bulk topologies as different theories.  
Hence, to take an appropriate measure for the quantum gravity, 
we expect
\begin{\eq}
\label{eq:eomegatopsum}
\mathcal{D}e \mathcal{D}\omega  
= \sum_{\rm bulk \, topology} \mathcal{D}\mathcal{A} \mathcal{D}\bar{\mathcal{A}} \,, 
\end{\eq}
and need to consider possible bulk topologies contributing to the partition function.
However our localization argument gives the answer to this sum; the bulk topology sum in the right hand side of 
\eqref{eq:eomegatopsum} is given by the modular sum (Rademacher sum) with {\it fixed} topology $D^2 \times S^1$, a solid torus \cite{Iizuka:2015jma}. 

To see this, let us recall that 
when we perform the localization calculation,  
configurations which contribute to the path integral are only around the localization locus. 
In this case, 
the metric and spin connection path integral $\mathcal{D}e \mathcal{D}\omega $ is replaced as $\mathcal{D}\mathcal{A} \mathcal{D}\bar{\mathcal{A}}$, but only the locus $\mathcal{F_{\mu\nu}}=0$ contributes to the path-integral, and 
$\mathcal{F_{\mu\nu}}=0$ is the same as the classical equation motion of the higher spin gravity including the Einstein equation! 
Since all the known classical solutions of the higher spin gravity have the topology $D^2 \times S^1$,  a solid torus 
(see e.g.~\cite{Castro:2011iw,Ammon:2012wc} for the known solutions), 
this would justify our claim that 
the partition function has contribution only from the bulk topology $D^2 \times S^1$. 
However this does not fix the bulk geometry for the gravity path integral completely. 
This is because the boundary of $D^2 \times S^1$ has topology of torus parametrized by complex structure $\tau$. 
One can recall that both $\tau$ and $-1/\tau$ gives the same asymptotic 
AdS boundary condition with the solid torus, $D^2 \times S^1$ bulk topology, 
but one corresponds to the thermal AdS while
the other corresponds to the non-rotating BTZ black hole \cite{Banados:1992wn}.
Although they are physically different,  
both of them have asymptotic AdS metric. 
Therefore we need to sum over all the consistent choices for the complex structure $\tau$. 
This is simply because  
in the localization calculation, we have to sum over all the possible configurations 
of the localization locus. 
For that, we ask which circle on the boundary torus is contractible and this  
information is labeled by cosets of $SL(2,\mathbb{Z})$,
which is parametrized by the integers $(a, b, c, d)$ for the modular transformation 
$\tau \rightarrow \frac{a\tau +b}{c\tau +d}$,  
satisfying $ad-bc=1$. 
Here $(a,b,c,d) \approx -(a,b,c,d)$ and $(c,d)_{\rm GCD} =1$, so that 
given $c \ge 0$ and $d$ satisfying $(c,d)_{\rm GCD} =1$, 
then $a$ and $b$ are uniquely determined up to the irrelevant ambiguity 
$(a, b) \approx (a + c, b + d)$. 
As a result, we are left with the sum over the integers $c \ge 0$ and $d$ and this gives the 
modular sum (Rademacher sum). Physically this is `roughly' equivalent to conducting the summation over all the 
bulk solutions as farey tail {\cite{Dijkgraaf:2000fq, Manschot:2007ha,Maloney:2007ud}}. 
But the localization gives more justification for this procedure.

Furthermore we assume holomorphic factorization of the partition function since the action \eqref{eq:action} 
is written as sum over holomorphic part $\mathcal{A}$ and anti-holomorphic part $\bar{\mathcal{A}}$ as \cite{Witten:2007kt}.
Then all of these considerations lead us to
\begin{\eq}
\mathcal{D}e \mathcal{D}\omega  
= \left( \sum_{c\geq 0,(c,d)_{\rm GCD}=1} \mathcal{D}\mathcal{A} \right) \left( \sum_{c\geq 0,(c,d)_{\rm GCD}=1} \mathcal{D}\bar{\mathcal{A}} \right) ,
\label{eq:measure}
\end{\eq} 
for the quantum gravity path integral, under the assumption of the localization at works.  
Thus the partition function takes the form
\begin{\eq}
Z_{\rm HigherSpin}  
= \Biggl| \sum_{c \geq 0,(c,d)_{\rm GCD}=1} \int \mathcal{D}\mathcal{A} \ e^{-\frac{ik}{4\pi} S_{\rm CS} [\mathcal{A}] } \Biggr|^2 .
\label{eq:wholeZ}
\end{\eq}
This factorization is actually reasonable in the localization method. 
This is because 
we regard $\mathcal{A}$ and $\bar{\mathcal{A}}$ as the independent 
variables and take the saddle point approximation for all of their locus. Since we need to sum 
over all the locus, the locus for $\mathcal{A}$ and $\bar{\mathcal{A}}$ need to be added independently as 
\eqref{eq:measure}. 

In the rest of this paper, 
we focus on the holomorphic part of the partition function 
and exactly compute this by the localization method.
Then we give a dual CFT interpretation for the holomorphic part.
As a conclusion we find that
the partition function has a natural interpretation \footnote{Dual CFTs for the pure $SL(N,\mathbb{C})$ 
higher spin gravity are proposed as 
minimal model by \cite{Castro:2011zq}  
for small values of the central charge, $c_Q < 2$.  
However, 
in this paper we do not consider such small central charge
because the level quantization condition forces us $c_Q \geq 24$.
} from 2D unitary CFT with $\mathcal{W}_N$-symmetry.

\section{Localization}
Now we compute the holomorphic part of the partition function \eqref{eq:wholeZ} of 
the CS theory on a space of the topology $D^2 \times S^1$ with appropriate boundary conditions.
Since $SL(N,\mathbb{C})$ is the complexification of $SU(N)$, and
for technical reason, we consider localization of $SU(N)$ Chern-Simons theory first, and then perform analytic continuation 
after the localization computation 
is done.
Now let us consider the following particular space of the topology $D^2 \times S^1$:
\begin{\eq}
ds^2 =d\theta^2 +\cos^2{\theta}(d\varphi^2 +\tan^2{\theta}dt_E^2 ),
\label{eq:space}
\end{\eq}
where $0\leq\theta\leq\theta_0 <\pi/2$, $0\leq \varphi \leq 2\pi$ and $0\leq t_E \leq 2\pi$.
Here, instead of the pure CS theory,
we supersymmetrize the CS theory on this space 
by introducing the 3D $\mathcal{N}=2$ vector multiplet $V=(\mathcal{A}_\mu ,\sigma ,D,\lambda ,\bar{\lambda })$ 
and compute its partition function by 
the localization technique \cite{Sugishita:2013jca}:
\begin{\eqa}
&&S_{\rm SCS}[V]
=S_{\rm CS} [\mathcal{A}] 
+\int d^3 x \sqrt{g} \, \tr \left( -\bar{\lambda}\lambda +2D\sigma \right) ,\NN\\
&&\int \mathcal{D}\mathcal{A} \ e^{-\frac{ik}{4\pi} S_{\rm CS} [\mathcal{A}] } \ \rightarrow\ 
\int \mathcal{D}V \ e^{-\frac{ik}{4\pi} S_{\rm SCS} [V] } \equiv Z_{c,d} \,, \qquad
\end{\eqa}
where $Z_{c,d}$ is the holomorphic partition function in \eqref{eq:wholeZ} of the supersymmetric CS theory 
with fixed choice for $(c,d)$. 
In gravity, the appropriate boundary condition for the metric is the choice 
giving 
asymptotic AdS boundary condition, 
which is Dirichlet boundary condition, rather than Neumann boundary condition \footnote{The CS theory with Neumann boundary condition is studied in \cite{Yoshida:2014ssa}.}.  
Then we can take the following boundary condition, 
keeping supersymmetry, 
\begin{\eqa}
&&\left. \mathcal{A}_\varphi \right|_{\theta =\theta_0} = a_\varphi ,\quad
\left. \mathcal{A}_{t_E} \right|_{\theta =\theta_0} = a_{t_E} ,\quad
\left. \sigma \right|_{\theta =\theta_0} = 0,\NN\\
&&\qquad \qquad \left.  \lambda \right|_{\theta =\theta_0}
=\left.  e^{-i(\varphi -t_E)} \gamma^\theta \bar{\lambda} \right|_{\theta =\theta_0} , 
\label{eq:bc}
\end{\eqa}
where all of them are proportional to the Cartan of $SU(N)$.
As a deformation term of the localization,
we choose $\mathcal{N}=2$ supersymmetric Yang-Mills term on the geometry \eqref{eq:space}: 
\begin{\eqa}
&& tS_{\rm YM} = t \int_{\mathcal{M}} {\rm Tr} \Bigl(  
\frac{1}{4} \mathcal{F}_{\mu\nu}^2 +\frac{1}{2} (D_\mu \sigma )^2 
+\frac{1}{2} \left( D+\frac{\sigma}{l} \right)^2 
\NN \\
&& \quad +\frac{i}{4} \bar{\lambda} \gamma^\mu D_\mu \lambda +\frac{i}{4}\lambda\gamma^\mu D_\mu \bar{\lambda}
     +\frac{i}{2}\bar{\lambda} [\sigma ,\lambda ] -\frac{1}{4l}\bar{\lambda} \lambda 
\Bigr) . \qquad
\end{\eqa}
Then localization locus is given by the flat connection $\mathcal{F}_{\mu\nu}=0$,
which is equivalent to the classical equation motion of the higher spin gravity including the Einstein equation. 
This is the reason why our topology sum is only for the $D^2 \times S^1$ in \eqref{eq:eomegatopsum}, \eqref{eq:measure}
and \eqref{eq:wholeZ}.

Since we have added just auxiliary fields,
one might think that this does not affect final result at first sight.
However we propose that this gives the following two effects.
First,
the CS level is renormalized as $k\rightarrow \keff =k+2$.
While the CS level shift generally occurs, 
we lack of a priori explanation that the renormalization for $k$ by two shift.
We will see later that 
our result has a natural interpretation from a dual CFT with the central charge $\keff N(N^2 -1)$,
which matches with the correct central charge extracted from the asymptotic symmetry analysis \cite{Henneaux:2010xg,Campoleoni:2010zq}
if the correct renormalization is $k\rightarrow \keff$. From this observation, we claim that 
the CS level is renormalized as $k\rightarrow \keff =k+2$. 
Second, 
there are extra $(N-1)$ massless fermionic degrees of freedom localized at the boundary only in the classical limit $k\rightarrow\infty$.
The boundary condition \eqref{eq:bc} kills the gaugino mass term $\bar{\lambda}\lambda$ at the boundary and
this implies that the $(N-1)$ boundary localized fermions appear in the classical limit $k\rightarrow\infty$.
The origin of $(N-1)$ fermions could be associated with the number of Cartan generators of the $SU(N)$.
This is because these Cartan elements of the fermion commute with the gauge field, $a_\mu$ in \eqref{eq:bc}. 
By using the doubling trick,
it is easy to see that classical solution for the fermion admits 
the boundary localized fermion wave function \cite{Iizuka:2015jma, Honda:2015hfa}.
Although it is a priori unclear if the boundary fermions exist also for finite $k$,
there is apparently no reason to expect the existence of the boundary fermions for finite $k$. 
Indeed the result of \cite{Iizuka:2015jma} for $N=2$ shows that $Z_{\rm hol}$ for $\keff =4$ 
is equal to conjectural $J$-function by Witten \cite{Witten:2007kt}, 
which agrees with the extremal CFT partition function of Frenkel, Lepowsky and Meurman \cite{FLM}.
This fact suggests that there is no boundary fermions for $\keff =4$.
Thus we propose that the boundary fermions exist only in the large $k$ limit. 

By using the results of \cite{Sugishita:2013jca,Iizuka:2015jma},
we obtain $Z_{c,d}$ as
\begin{\eq}
Z_{c,d} =e^{ik\pi {\rm tr}(a_\varphi a_{t_E})} 
 \prod_{\alpha \in {\rm root}_+} \left( e^{ i\pi\alpha (a_\varphi )} -e^{- i\pi\alpha (a_\varphi )}  \right) ,
\label{eq:Zcd1} 
\end{\eq}
where $\alpha$ runs the positive root of $SU(N)$ and $a_\varphi$ and $a_{t_E}$ take appropriate values as explained below.
The first and second terms come from the classical contribution \cite{Banados:2012ue} and the one-loop determinant in the localization, respectively.

In order to compute $Z_{c,d}$,
it is sufficient to compute the case of non-rotating BTZ black hole \cite{Banados:1992wn,Ammon:2012wc}, 
namely $(c,d)=(1,0)$. 
Since BTZ black holes are solutions for the $SL(2,\mathbb{C})$ case and 
we are considering generic $SL(N,\mathbb{C})$ case, 
how we describe 
BTZ black holes in $SL(N,\mathbb{C})$ depends on how we embed the $SL(2,\mathbb{C})$ sector into the whole $SL(N,\mathbb{C})$. As we claim before, here 
we choose principal embedding, and there 
$L_0$ is given by
\begin{\eq}
\label{eq:L0repn}
L_0 ={\rm diag}(\lambda_1 ,\cdots ,\lambda_N )
\, \,
{\rm with}\ \lambda_m =\frac{1}{2}(N+1-2m) \,.
\end{\eq}

Given this, let us recall 
the boundary condition for the holonomy 
in the case of 
the
non-rotating BTZ black hole. 
In $N=2$ case, it is given by \cite{Iizuka:2015jma}:
\begin{\eq}
\label{eq:boudaryholoN2}
a_\varphi =\frac{1}{\tau} L_0 \,,\quad a_{t_E} = L_0 \,, 
\end{\eq}
where $L_0$ is given by $N=2$ case of \eqref{eq:L0repn}.

Since BTZ black hole solutions without higher spin charges in the higher spin gravity theory are just embedding of the $SL(2,\mathbb{C})$ sector into the whole $SL(N,\mathbb{C})$,
the condition \eqref{eq:boudaryholoN2} is still true for generic $N$ 
if $L_0$ is replaced by  
$N \times N$ matrix representation in $SL(N,\mathbb{C})$ given by  \eqref{eq:L0repn}.  
To see this more explicitly, note that in the higher spin gravity, a BTZ black hole is written as 
(see \cite{Ammon:2012wc}, but take care the difference for the convention $i$ 
and periodicity for $t_E$)
\begin{\eq}
\label{AzboundaryconditionforzerochargeBH}
A(z)= \frac{1}{i}  \Big( L_1 -{2\pi \over k} {\cal{L}}(z) L_{-1} \Big)  \left(d \varphi + \tau d t_E \right)
\,,
\end{\eq}
where $L_1$, $L_{-1}$ are $N\times N$ matrix principal embedding. More explicitly, see appendix A in \cite{Castro:2011iw}. 
Diagonalizing this matrix by gauge transformation, and using the smoothness condition 
at the horizon, namely eq.~(2.20) of \cite{Ammon:2012wc}, we obtain 
\begin{\eq}
\label{draftholo}
A(z) =  \frac{1}{\tau}  L_0  \left(d \varphi + \tau d t_E \right) \,, 
\end{\eq}
which is exactly the same form as \eqref{eq:boudaryholoN2}, though $L_{0}$ is given instead by $N \times N$ matrix \eqref{eq:L0repn}. 
Therefore 
we obtain, for generic $N$ case, 
\begin{\eq}
 a_\varphi =\frac{1}{\tau} {\rm diag}(\lambda_1 ,\cdots ,\lambda_N ) \, \,, 
 \,
a_{t_E} = {\rm diag}(\lambda_1 ,\cdots ,\lambda_N ) \,. 
\label{a10}
\end{\eq}
Plugging this into \eqref{eq:Zcd1}, we find {(See appendix for more detail.)}
\begin{\eq}
Z_{1,0} 
=e^{\frac{2\pi i}{\tau}\frac{k+2}{24}N(N^2 -1) } 
 \prod_{s=2}^N\prod_{l=1}^{s-1} \left( 1 -e^{-\frac{2\pi i l}{\tau}}  \right) \,.
 \label{Z10}
\end{\eq}
Then $Z_{c,d}$ can be obtained by the modular transformation of $Z_{1,0}$ as
$Z_{c,d} =\left. Z_{1,0} \right|_{-\frac{1}{\tau}\rightarrow \frac{a\tau +b}{c\tau +d} }$. 
This leads us to
\begin{\eqa}
Z_{c,d} 
&=& e^{- 2 \pi i \frac{k+2}{24}N(N^2 -1) \frac{a\tau +b}{c\tau +d} } 
 \prod_{s=2}^N\prod_{l=1}^{s-1} \left( 1 -e^{2 \pi il \frac{a\tau +b}{c\tau +d}}   \right)  \NN\\
&=& e^{-2 \pi i \frac{k+2}{24}N(N^2 -1) \frac{a\tau +b}{c\tau +d} } 
\sum_{m=0}^{\frac{1}{6}N(N^2 -1)} a_m e^{2 \pi i m \frac{a\tau +b}{c\tau +d}} \,, \qquad 
\label{eq:Zcd}
\end{\eqa}
where $a_m$ is defined by
\begin{\eq}
\label{eq:anidentity}
\sum_{m=0}^{\frac{1}{6}N(N^2 -1)} a_m x^m
=\prod_{s=2}^N\prod_{l=1}^{s-1} \left( 1 -x^l  \right) \,.
\end{\eq}

Next we perform the summation over $(c,d)_{\rm GCD}$ in \eqref{eq:wholeZ}.
We regularize this summation by Rademacher sum \cite{rademacher:1939} as in the $N=2$ case \cite{Iizuka:2015jma}
and obtain the following holomorphic partition function for the pure higher spin gravity
\begin{\eqa}
Z_{\rm hol}(q)
&=& Z_{0,1}(\tau ) +\sum_{c>0, (c,d)_{\rm GCD}=1} \left( Z_{c,d}(\tau ) -Z_{c,d}(\infty ) \right) \NN\\
&=& \sum_{m=0}^{\frac{1}{6}N(N^2 -1)} a_m 
R^{( - \frac{ N(N^2 -1) \keff }{24} + m )}(q) \,,
\label{eq:Zhol} 
\end{\eqa} 
where $\keff =k+2$ and $q = e^{2 \pi i \tau}$, and 
\begin{\eq} 
R^{(m)}(q) = e^{2\pi im\tau} 
+\sum_{c>0, (c,d)_{\rm GCD}=1} \left( e^{2\pi im\frac{a\tau +b}{c\tau +d}} -e^{2\pi im\frac{a}{c}}\right) .
\end{\eq}
This is one of the main results in this paper. 
For example, explicit results for $N=2, 3, 4, \cdots$ are
\begin{\eqa}
\left. Z_{\rm hol} (q) \right|_{N=2}
&=&  R^{( -\frac{\keff}{4} )}(q) - R^{( - \frac{\keff}{4} +1 )}(q) \,, \NN\\
\left. Z_{\rm hol} (q) \right|_{N=3}
&=& R^{(-\keff )} -2R^{(-\keff +1)} \NN\\
&& +2 R^{(-\keff +3)} - R^{(-\keff +4)} \,, \NN \\
\left. Z_{\rm hol} (q) \right|_{N=4}
&=& R^{( - \frac{5 \keff}{2} )} - 3 R^{( -\frac{5 \keff}{2} +1 )}
+ R^{( - \frac{5 \keff}{2} +2)} \NN\\
&& \, + 4 R^{( - \frac{5 \keff}{2} +3)} - 2 R^{( - \frac{5 \keff}{2} +4)} \NN\\
&& \,\, - 2 R^{ - \frac{5 \keff}{2} +5 )} - 2 R^{( - \frac{5 \keff}{2} +6)} \NN\\
&& \,\,\, + 4 R^{( - \frac{5 \keff}{2} +7)} + R^{( - \frac{5 \keff}{2} +8)}  \NN\\
&& \,\,\,\, - 3 R^{( - \frac{5 \keff}{2} +9)} + R^{( - \frac{5 \keff}{2} +10)} \,,
\label{explicitZhol}
\end{\eqa}
and so on. It is straightforward to obtain explicit form for $N \ge 5$, and  
$N=2$ case is the one obtained in \cite{Iizuka:2015jma}.

As we have discussed in \cite{Iizuka:2015jma}, for $m < 0$ in $R^{(m)}(q)$, $m$ needs to be 
integer; 
This is because $R^{(m)}(q) = q^m + (\mbox{non-singular \, terms})$ in $q \to 0$ limit for $m < 0$. 
Therefore 
$q^m$ dominates in the RHS, and then the modular invariance requires that $m$ should be integer. 
See also Proposition 5.4.2. in \cite{Duncan:2009sq}. 
Note that 
the quantization condition $k_{eff}/4 \in\mathbb{Z}$ guarantees $c_Q/24 \in \mathbb{Z}$ with $c_Q =\keff N(N^2 -1)$.   
Therefore, this guarantees that  $ - \frac{ N(N^2 -1) \keff }{24} + m $, the ``orders''  for $R^{( - \frac{ N(N^2 -1) \keff }{24} + m )}(q)$ in 
\eqref{eq:Zhol} and \eqref{explicitZhol} are always integer.

\section{Dual CFT interpretation for large $k$}
As discussed above, 
$(N-1)$ boundary localized fermions are expected to exist in the large $k$ limit, 
and a contribution from each fermion 
to the partition function
is expected to be 
\begin{\eq}
Z_{\rm B-fermion} = \prod_{n =1}^\infty (1-q^n ) \,.
\end{\eq}
as is discussed in \cite{Iizuka:2015jma, Honda:2015hfa}. 
Since these boundary fermions are decoupled in the large $k$ limit,
we claim that 
(holomorphic part of) `bulk pure higher spin gravity' partition function is given by
\begin{\eq}
Z_{\rm HigherSpin}^{ {\rm large}\ k} (q)
= \frac{Z_{\rm hol} (q) }{(Z_{\rm B-fermion})^{N-1}} \,.
\end{\eq}
In this section we discuss properties of $Z_{\rm HigherSpin}^{ {\rm large}\ k} (q)$ and its dual CFT interpretation.
For this purpose it is convenient to use the following representation of the Rademacher sum $R^{(m)} (q)$ 
\begin{\eq}
R^{(m)} (q) =q^m +({\rm const.}) +\sum_{n=1}^\infty c(m,n) q^n \,,
\label{eq:R_smallq}
\end{\eq}
where
\begin{\eqa}
c(m,n)
&=& 2\pi\sqrt{\frac{-m}{n}}
\sum_{c=1}^\infty \frac{A_c (m,n)}{c} I_1 \left( \frac{4\pi\sqrt{-mn}}{c} \right)  \,,\NN\\
A_c (m,n)
&=& \sum_{1\leq d \leq c , (c,d)_{\rm GCD}=1} e^{2\pi i (m\frac{a}{c} +n\frac{d}{c})} \,.
\label{eq:coeffs}
\end{\eqa}
Here the sum for $A_c (m,n)$ is only for $d$ and 
$I_1 (z)$ is the modified Bessel function of the first kind.
Since the (const) term is regularization dependent,
here we simply set the (const) term to be zero for convenience.
By plugging the expression \eqref{eq:R_smallq} into \eqref{eq:Zhol}, and by 
using \eqref{eq:anidentity} and the identity 
\begin{\eq}
\prod_{s=2}^N\prod_{l=1}^{s-1} \left( 1 - q^l  \right) = \frac{\prod_{n =1}^\infty (1-q^n )^{N-1}}{\prod_{s=2}^N\prod_{m=s}^{\infty} \left( 1 - q^m  \right) } \,, 
\end{\eq}
we find 
\begin{\eqa}
Z_{\rm hol}(q)
&=& (Z_{\rm B-fermion} )^{N-1} q^{-\frac{c_Q}{24} } \prod_{s=2}^N \prod_{m=s}^\infty \frac{1}{1-q^m}  \NN\\
&& +\sum_{\Delta  =1}^\infty c_\Delta^{(\keff )} q^\Delta ,
\label{eq:leading}
\end{\eqa}
where $c_Q =\keff N(N^2 -1)$ and
\begin{\eq}
c_\Delta^{(\keff )}
=\sum_{m=0}^{\frac{1}{6}N(N^2 -1)} a_m \, c\left( -\frac{c_Q}{24} + m , \Delta \right) \,.
\end{\eq}
Then $Z_{\rm HigerSpin}^{{\rm large}\ k}$ becomes
\begin{\eqa}
Z_{\rm HigerSpin}^{{\rm large}\ k}(q)
&=& q^{-\frac{c_Q}{24} } \prod_{s=2}^N \prod_{m=s}^\infty \frac{1}{1-q^m}  \NN\\
&& +\sum_{\Delta  =1}^\infty c_\Delta^{(\keff )} \frac{q^\Delta}{\prod_{n=1}^\infty (1-q^n )^{N-1}} \,. \,\,
\label{eq:Verma}
\end{\eqa}
Note that the first factor is the same as the character of the Verma module
constructed from the vacuum $|0\rangle$ in $\mathcal{W}_N$ CFT (see e.g. sec.~6.3.2 of \cite{Bouwknegt:1992wg}) 
\begin{\eq}
Z_{\rm vac}(q) 
= {\rm Tr}_{V|0\rangle} q^{L_0 -\frac{c_Q}{24}}
=q^{-\frac{c_Q}{24} } \prod_{s=2}^N \prod_{m=s}^\infty \frac{1}{1-q^m} \,,
\end{\eq}
where the vacuum satisfies 
\begin{\eq}
W^{(s)}_{n\geq -(s-1)} |0\rangle =0 \, \quad \mbox{(for $s=2,\cdots ,N$)} \,. 
\end{\eq}
$s$ represents spin  
and $W_n^{(s)}$ is the generator of the $\mathcal{W}_N$-algebra.
In particular, $W_n^{(s=2)}$ is the Virasoro generator $L_n$. 
From this vacuum, we can construct eigenstates of $L_0$ by
\begin{\eq}
\prod_{s=2}^N \prod_{m^{(s)} =s}^\infty ( W^{(s)}_{-m^{(s)}} )^{\ell_{m^{(s)}}}   |0\rangle ,
\end{\eq}
with the eigenvalue
$-c_Q/24 +\sum_{s=2}^N \sum_{m^{(s)}=s}^\infty m^{(s)} \ell_{m^{(s)}}  $.
Note also that
the vacuum character takes the same form as
(holomorphic part of) the one-loop partition function of the higher spin gravity on $AdS_3$ \cite{Gaberdiel:2010ar}.
Similarly, the last factor in \eqref{eq:Verma} is the same as
the character $Z_{\rm primary}^{(\Delta )}$ of the Verma module 
generated by the state $|\Delta \rangle $ satisfying
$W^{(s)}_{n>0} |\Delta \rangle =0$ and $L_0 |\Delta \rangle =(c_Q /24 +\Delta )|\Delta \rangle$ 
in $\mathcal{W}_N$-CFT
\begin{\eq}
Z_{\rm primary}^{(\Delta )} (q)
= {\rm Tr}_{V|\Delta \rangle} 
 q^{L_0 -\frac{c_Q}{24}} 
= \frac{ q^\Delta}{{\prod_{m=1}^\infty (1-q^m )^{N-1}}} \,. 
\end{\eq}
Thus we find the following nice expansion for the partition function 
in terms of the characters of the vacuum and primaries,
\begin{\eq}
Z_{\rm HigherSpin}^{{\rm large}\ k} (q) 
= Z_{\rm vac}(q)
+ \sum_{\Delta =1}^\infty c_\Delta^{(\keff )} Z_{\rm primary}^{(\Delta )}  (q) \,.
\end{\eq}
Furthermore we also claim that
$c_\Delta^{(\keff )}$ is positive integer 
as expected if the dual CFT is unitary.
We have explicitly checked this 
for $(N;\keff ;\Delta )=(2,\cdots ,14; 4,\cdots ,k_{\rm max}(N); 1,\cdots ,50)$,
with $( k_{\rm max}(2) ,\cdots ,k_{\rm max}(14))$ $=$ $(2000$, $500$, $200$, $100$, $56$, $32$, $20$, $16$, $12$, $8$, $4$, $4$, $4)$.
For example, we have
\begin{\eqa}
&&\left. c_{\Delta= 1}^{(4)} \right|_{N=2} = 196884 ,\quad
\left. c_{\Delta= 2}^{(4)} \right|_{N=2} = 21493760,\NN\\
&&\left. c_{\Delta= 1}^{(4)} \right|_{N=3} = 75798018972,\NN\\
&&\left. c_{\Delta= 2}^{(4)} \right|_{N=3} = 1580451492798464,\NN\\
&&\left. c_{\Delta= 1}^{(4)} \right|_{N=4} = 144317861960158148.
\end{\eqa}
\begin{figure}[t]
\begin{center}
\includegraphics[width=8.8cm]{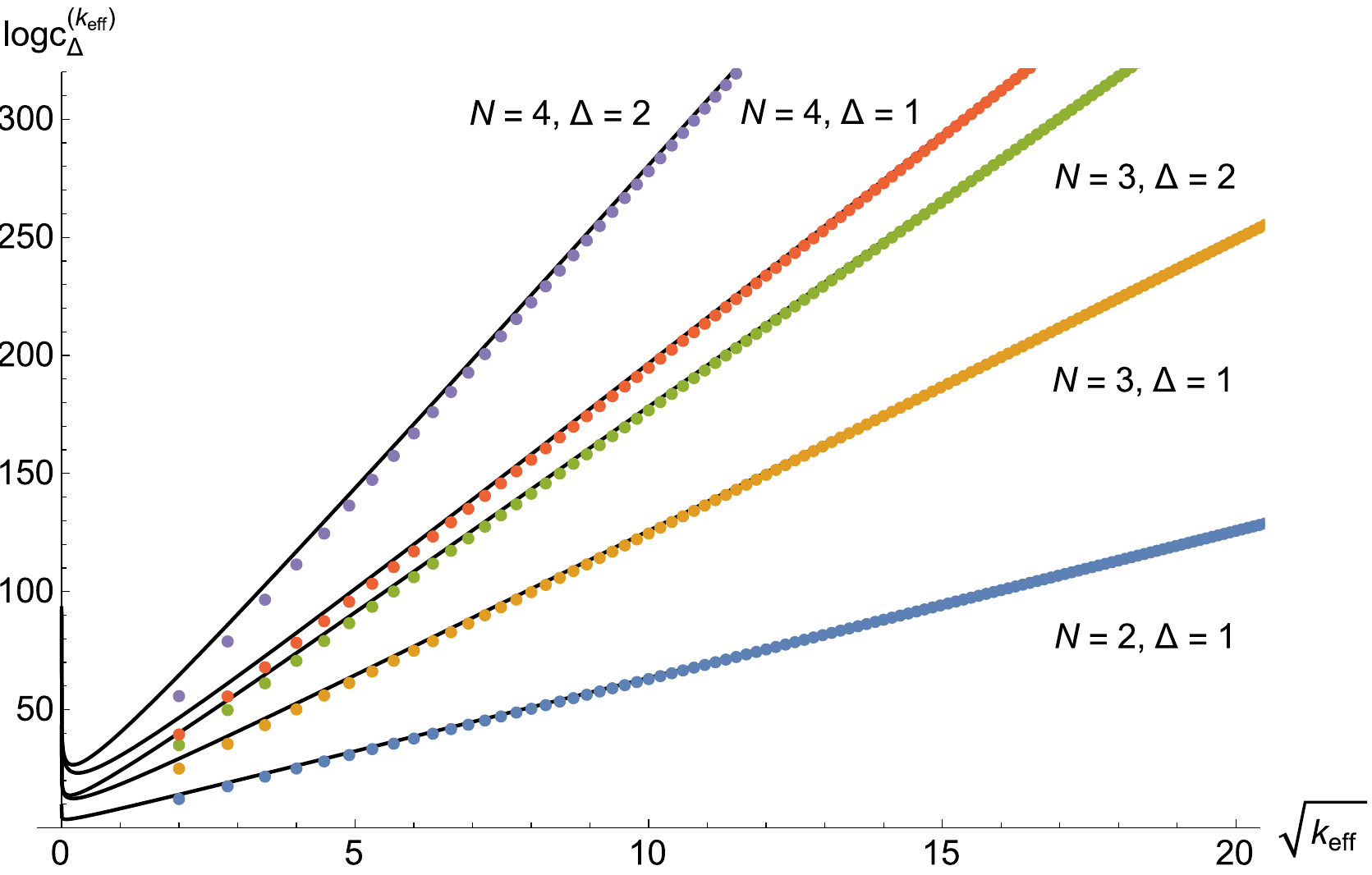}
\end{center}
\caption{
The log of the $\mathcal{O}(q^\Delta )$ coefficient $c_\Delta^{(\keff )}$ is plotted to $\sqrt{\keff}$
for $(N,\Delta )=(2,1),(3,1),(3,2),(4,1)$ and $(4,2)$.
The solid lines denote the Cardy formula 
including the sub-leading corrections,
which is explicitly written in (37). 
}
\label{fig:entropy}
\end{figure}
We can also see that
the $\mathcal{O}(q^\Delta )$ coefficient $c_\Delta^{(\keff )}$ exhibits the Cardy formula in the large $k$ limit.
Since the modified Bessel function asymptotically behaves as 
\begin{\eq}
\lim_{z\rightarrow\infty} I_\nu (z) 
\sim \frac{e^z}{\sqrt{2\pi z}}\ _2 F_0 \left( \nu +\frac{1}{2}, \frac{1}{2}-\nu ;\frac{1}{2z}\right) ,
\end{\eq}
in the summation \eqref{eq:coeffs} for $c( m,n )$,
the $c=1$ term gives most dominant contribution in the large $k$ limit.
Noting $A_1 (m,n)=1$ for $^\forall m,n\in\mathbb{Z}$, we find
\begin{\eqa}
c(m,n)
&\sim & \frac{(-m)^{1/4}}{\sqrt{2}n^{3/4}} e^{4\pi\sqrt{-mn}}\ _2 F_0 \left( \frac{3}{2}, -\frac{1}{2} ;\frac{1}{8\pi\sqrt{-mn}}\right) \NN\\
&&+\mathcal{O}(e^{2\pi\sqrt{-mn}}) .
\end{\eqa}
This immediately shows that 
the coefficients $c_\Delta^{(\keff )}$ of the characters asymptotically behave as
\begin{\eq}
\lim_{k\rightarrow\infty} \log{c_\Delta^{(\keff )}}
=2\pi\sqrt{\frac{c_Q \Delta}{6}} \,. 
\end{\eq}
This perfectly matches with the Cardy formula for 2D CFT in the large central charge limit $c_Q \rightarrow\infty$, 
which is the same as the large $k$ limit, and therefore 
this result agrees with the entropy of higher spin black hole \cite{Ammon:2012wc}.
It is also easy to compute 
sub-leading corrections to the Cardy formula.  
Taking into account the leading and sub-leading orders, we find, for example,
\begin{\eqa}
\lim_{k\rightarrow\infty} \left. c_\Delta^{(\keff )} \right|_{N=2} 
&= & 2\pi\sqrt{\frac{c_Q \Delta}{6}} +\log{\Biggl( \frac{2\cdot 6^{\frac{1}{4}} \pi  }{\Delta^{\frac{1}{4}} c_Q^{\frac{1}{4}} } \Biggr)} \,,\NN\\
\lim_{k\rightarrow\infty} \left. c_\Delta^{(\keff )} \right|_{N=3} 
&=& 2\pi\sqrt{\frac{c_Q \Delta}{6}} +\log{\Biggl( \frac{384 \cdot 6^{\frac{1}{4}} \pi ^3  \Delta^{\frac{3}{4}}}{c_Q^{\frac{5}{4}}} \Biggr)} \,,\NN\\
\lim_{k\rightarrow\infty} \left. c_\Delta^{(\keff )} \right|_{N=4} 
&=& 2\pi\sqrt{\frac{c_Q \Delta}{6}} +\log{\Biggl( \frac{884736\cdot 6^{\frac{3}{4}} \pi ^6  \Delta^{\frac{9}{4}}}{c_Q^{\frac{11}{4}}} \Biggr)} \,. \NN\\
\label{eq:Cardysub}
\end{\eqa}
Similarly, computations of any higher order corrections to $c_\Delta^{(\keff)}$ in $\frac{1}{\sqrt{c_Q}}$ expansion is straightforward. 

These formula give good approximation for precise value of $c_\Delta^{(\keff )}$. 
In FIG.~\ref{fig:entropy},
we plot $\log{c_\Delta^{(\keff )}}$ (dotted lines) as a function of $\sqrt{\keff}$ 
and compare those with the Cardy formula with the sub-leading corrections \eqref{eq:Cardysub} (solid lines). 
One can easily see that the solid lines  
match with 
the dotted lines  
even for large but finite $k$, and that 
the coefficients approach to 
the behavior \eqref{eq:Cardysub} perfectly 
in the large $k$ regime.

\section{Summary and Discussion}
In this paper we have studied the partition function of the 3D higher spin gravity with negative cosmological constant
by generalizing the previous works \cite{Iizuka:2015jma,Honda:2015hfa} 
on the pure Einstein gravity.
We first rewrite the higher spin gravity in 
the supersymmetric way and
exactly compute the partition function of this theory by using the localization method. 
The final answer of the partition function is written in 
a nice way 
in terms of the characters for the vacuum and primaries in 2D CFT with $\mathcal{W}_N$-symmetry. 
We also check that the coefficients of the character expansion are positive integers 
and exhibit the Cardy formula in the large central charge limit.

In our procedure, we have assumed that 
the supersymmetrization does not essentially change the partition function
but gives the following two effects: 
First,
we assume that the CS level is renormalized as $k\rightarrow \keff =k+2$.
Although the CS level shift for compact gauge group is 
usually dependent on the rank of the gauge group or zero,
we claim that the renormalization is nonzero but independent of $N$.
As a result, our final result has 
the natural interpretation from the dual CFT with the central charge $\keff N(N^2 -1)$,
which matches with the central charge extracted from the asymptotic symmetry analysis \cite{Henneaux:2010xg,Campoleoni:2010zq}.
It is interesting if one can justify this by some arguments.
Second, 
we assume that there are the extra $(N-1)$ fermions localized at the boundary only in the classical limit $k\rightarrow\infty$.
There is apparently no reason to expect the existence of the boundary fermions also for finite $k$. 
Indeed the result of \cite{Iizuka:2015jma} for $N=2$ shows that $Z_{\rm hol}$ for $\keff =4$ 
is equal to conjectural $J$-function by Witten \cite{Witten:2007kt}
and suggests that there is no boundary fermions for $\keff =4$.
Thus we propose that the boundary fermions exist only in the large $k$ limit.

We have also assumed that
the summation over the topologies in the quantum higher spin gravity
has the contributions only from the topology of the solid torus. 
Our localization argument shows that 
the higher spin gravity partition function is given by configurations around the localization locus $\mathcal{F}_{\mu\nu}=0$,
which is the same as the classical equation motion of the higher spin gravity including the Einstein equation. 
Since all 
the
known solutions of the 3D Einstein equation have the topology of the solid torus,
this would justify our assumption.

Note that all of the localization locus $\mathcal{F}_{\mu\nu}=0$ which contribute to our final expression for the partition  
function are BTZ black holes.  
In other words, since the partition function \eqref{eq:higherspinpartitionZ} corresponds to the {\it zero} chemical potential for higher spin charge, 
our results are expressed in terms of summing over higher-spin-charge neutral black holes,  
and small fluctuations around those 
are
characterized by $\mathcal{W}_N$ algebra. 
These small fluctuations give exactly the 
one-loop partition function of the higher spin gravity on $AdS_3$ \cite{Gaberdiel:2010ar} 
and contains nonzero higher-spin-charge states. 
The appearance of the $\mathcal{W}_N$-symmetry is natural
since this is the same as the asymptotic symmetry of the higher spin gravity \cite{Campoleoni:2010zq,Campoleoni:2011hg}.
The coefficients $c_\Delta^{(\keff )}$ of the character expansion are positive integer and
exhibit the Cardy formula in the large $k$ limit.

It is interesting to introduce nonzero higher-spin-charge chemical potentials. Then, 
the localization locus contributing to the final answer are black holes with nonzero higher-spin-charges. 
For example, let us consider introducing a chemical potential for the spin-3-charge. The effects of introducing such a chemical potential is reflected as the change of the boundary condition for the gauge fields. More concretely, eq.~\eqref{AzboundaryconditionforzerochargeBH}, which we use for the case of zero chemical potential, must be replaced by eq.~(3.13) of \cite{Ammon:2012wc} for the case with nonzero spin-3-charge chemical potential \footnote{Again we have to take care the difference for the convention $i$ and periodicity for $t_E$, see the difference between eq.~\eqref{AzboundaryconditionforzerochargeBH} and eq.~(3.7) of the reference \cite{Ammon:2012wc}.}, where  $\cal{L}$ and $\cal{W}$ in eq.~(3.13) are related to the spin-3-charge chemical potential $\mu$ and temperature $\tau$ by eq.~(3.25) of that reference.  As we diagonalize all the components of these boundary condition from eq.~\eqref{AzboundaryconditionforzerochargeBH} to eq.~\eqref{draftholo}, we now need to diagonalize the new boundary condition, which corresponds to eq.~(3.13) of \cite{Ammon:2012wc}. Note that due to the commutativity of $[ a_z , a_{\bar z} ] = 0$ as shown in eq.~(3.14) of \cite{Ammon:2012wc}, it is possible to diagonalize all the components of these boundary gauge fields. Once we obtain the diagonalized values, which should replace our eq.~\eqref{a10}, then the rest of the calculation is straightforward and we can obtain the final partition function.  
However we will not proceed the rest of calculation here.

In this paper, we choose the principal embedding but it would also be interesting to study non-principal embedding and 
how the results are modified by that. 
Finally but no lastly, it is interesting to generalize our analysis to supergravity.

\section*{APPENDIX}
{
Here, we derive the formula \eqref{Z10} from \eqref{eq:Zcd1}.
For $(c,d)=(1,0)$, we take $a_\varphi, a_{t_E}$ as \eqref{a10},
and 
\begin{align}
Z_{1,0}
&=
e^{ \frac{ik \pi}{\tau} \sum_{m=1}^N \lambda_m^2}
\prod_{1\leq m < n \leq N}
\Big(
e^{ \frac{i \pi}{\tau} (\lambda_m - \lambda_n)  }
-
e^{-\frac{i \pi }{\tau} (\lambda_m - \lambda_n)  }
\Big)
\notag \\
&=
e^{ \frac{2\pi i}{\tau} \frac{k}{24} N(N^2-1) }
\prod_{1\leq m < n \leq N}
e^{\frac{\pi i (n-m)}{\tau} }
\Big(
1 
-
e^{-\frac{2 \pi i (n-m)}{\tau}  }
\Big)
\notag \\
&=
e^{ \frac{2\pi i}{\tau} \frac{k}{24} N(N^2-1) }
\prod_{l=1}^{N-1}
\prod_{m=1}^{N-l}
e^{\frac{\pi i l}{\tau} }
\Big(
1 
-
e^{-\frac{2 \pi i l}{\tau}  }
\Big)
\notag \\
&=
e^{ \frac{2\pi i}{\tau} \frac{k+2}{24} N(N^2-1) }
\prod_{l=1}^{N-1}
\Big(
1 
-
e^{-\frac{2 \pi i l}{\tau}  }
\Big)^{N-l} \,.
\end{align}
Here we change the product for $m$ and $n$ into $l$ and $m$, where $l= n-m $ from the second equality to the 
third equality. Finally, we rewrite the product as
\begin{align}
\label{eq:magicaleq}
\prod_{l=1}^{N-1}
\Big(
1 
-
e^{-\frac{2 \pi i l}{\tau}  }
\Big)^{N-l}
=
\prod_{s=2}^N
\prod_{l=1}^{s-1}
\Big(
1- e^{- \frac{2 \pi i l}{\tau}}
\Big) 
\,.
\end{align}
This is just re-ordering the product.
It can be understood by the following table.
\begin{align}
\left. \begin{array}{l||c|c|c|c|c|}
(s=2) &l=1& & & & \\
 \hline
(s=3) &l=2& l=1& & & \\
 \hline
(s=4) &l=3 & l=2 &l=1&& \\
 \hline
\dots &\dots&\dots&\dots&\dots& \\
 \hline
(s=N) & l=N-1 & l=N-2 & l=N-3 & \dots & l=1
\end{array} \right.
\notag
\end{align}
In this table, we arrange $(N-1)$ number of $l=1$ in the diagonal slots, and $(N-2)$ number of $l=2$ in second slots, etc.
\eqref{eq:magicaleq} can be understood from this table. LHS is products for diagonal elements and RHS is products for row elements.  
}

\vspace{.5cm}
\acknowledgments
AT would also like to thank Osaka University for its hospitality where part of this work was done. 
The work of NI was supported in part by 
JSPS KAKENHI Grant Number 25800143. 
The work of AT was supported in part by the RIKEN iTHES Project.



\begin{thebibliography}{99}

\bibitem{Iizuka:2015jma} 
  N.~Iizuka, A.~Tanaka and S.~Terashima,
  ``Exact Path Integral for 3D Quantum Gravity,''
  Phys.\ Rev.\ Lett.\  {\bf 115}, no. 16, 161304 (2015)
  [arXiv:1504.05991 [hep-th]].

\bibitem{Honda:2015hfa} 
  M.~Honda, N.~Iizuka, A.~Tanaka and S.~Terashima,
  ``Exact Path Integral for 3D Quantum Gravity II,''
  arXiv:1510.02142 [hep-th].

\bibitem{Achucarro:1987vz}
A.~Achucarro and P.~Townsend, 
``A Chern-Simons Action for Three-Dimensional
  anti-De Sitter Supergravity Theories,"  
Phys.Lett. {\bf B180} (1986) 89.

\bibitem{Witten:1988hc}
E.~Witten, ``(2+1)-Dimensional Gravity as an Exactly Soluble System,"
Nucl.Phys. {\bf B311} (1988) 46.



\bibitem{Blencowe:1988gj}
M.~Blencowe, 
``A Consistent Interacting Massless Higher Spin Field Theory in $D$ = (2+1),"
Class.Quant.Grav. {\bf 6} (1989) 443.

\bibitem{Vasiliev:2003ev} 
  M.~A.~Vasiliev,
  ``Nonlinear equations for symmetric massless higher spin fields in (A)dS(d),''
  Phys.\ Lett.\ B {\bf 567}, 139 (2003)
  [hep-th/0304049].

\bibitem{Klebanov:2002ja}
I.~R. Klebanov and A.~M. Polyakov, 
``AdS dual of the critical O(N) vector model,  
Phys. Lett. {\bf B550} (2002) 213--219,
hep-th/0210114.

\bibitem{Gross:1988ue}
D.~J. Gross, 
``High-Energy Symmetries of String Theory,"
Phys.Rev.Lett. {\bf 60} (1988) 1229.

\bibitem{Ammon:2012wc}
M.~Ammon, M.~Gutperle, P.~Kraus, and E.~Perlmutter, 
``Black holes in three dimensional higher spin gravity: A review,"
J. Phys. {\bf A46} (2013) 214001, 
arXiv:1208.5182.

\bibitem{Castro:2011iw} 
  A.~Castro, R.~Gopakumar, M.~Gutperle and J.~Raeymaekers,
  ``Conical Defects in Higher Spin Theories,''
  JHEP {\bf 1202}, 096 (2012)
  [arXiv:1111.3381 [hep-th]].

\bibitem{Campoleoni:2010zq}
A.~Campoleoni, S.~Fredenhagen, S.~Pfenninger, and S.~Theisen, 
``Asymptotic symmetries of three-dimensional gravity coupled to higher-spin fields,"
 JHEP {\bf 1011} (2010) 007, arXiv:1008.4744.


\bibitem{Campoleoni:2011hg}
A.~Campoleoni, S.~Fredenhagen, and S.~Pfenninger,
``Asymptotic W-symmetries
  in three-dimensional higher-spin gauge theories,"  
JHEP {\bf 09} (2011) 113, 
arXiv:1107.0290.

\bibitem{Brown:1986nw}
J.~D. Brown and M.~Henneaux, 
``Central Charges in the Canonical Realization
  of Asymptotic Symmetries: An Example from Three-Dimensional Gravity,"
 Commun. Math. Phys. {\bf 104} (1986) 207--226.

\bibitem{Henneaux:2010xg}
M.~Henneaux and S.-J. Rey, 
``Nonlinear $W_{infinity}$ as Asymptotic
  Symmetry of Three-Dimensional Higher Spin Anti-de Sitter Gravity,"
  JHEP {\bf 1012} (2010) 007,
  arXiv:1008.4579.

\bibitem{Witten:1989ip} 
  E.~Witten,
  ``Quantization of {Chern-Simons} Gauge Theory With Complex Gauge Group,''
  Commun.\ Math.\ Phys.\  {\bf 137}, 29 (1991).

\bibitem{Witten:2007kt} 
  E.~Witten,
  ``Three-Dimensional Gravity Revisited,''
  arXiv:0706.3359 [hep-th].

\bibitem{Banados:1992wn} 
  M.~Banados, C.~Teitelboim and J.~Zanelli,
  ``The Black hole in three-dimensional space-time,''
  Phys.\ Rev.\ Lett.\  {\bf 69}, 1849 (1992)
  [hep-th/9204099].


\bibitem{Dijkgraaf:2000fq} 
  R.~Dijkgraaf, J.~M.~Maldacena, G.~W.~Moore and E.~P.~Verlinde,
  ``A Black hole Farey tail,''
  hep-th/0005003.

\bibitem{Manschot:2007ha} 
  J.~Manschot and G.~W.~Moore,
  ``A Modern Farey Tail,''
  Commun.\ Num.\ Theor.\ Phys.\  {\bf 4}, 103 (2010)
  [arXiv:0712.0573 [hep-th]].

\bibitem{Maloney:2007ud} 
  A.~Maloney and E.~Witten,
  ``Quantum Gravity Partition Functions in Three Dimensions,''
  JHEP {\bf 1002}, 029 (2010)
  [arXiv:0712.0155 [hep-th]].

\bibitem{Castro:2011zq} 
  A.~Castro, M.~R.~Gaberdiel, T.~Hartman, A.~Maloney and R.~Volpato,
  ``The Gravity Dual of the Ising Model,''
  Phys.\ Rev.\ D {\bf 85}, 024032 (2012)
  [arXiv:1111.1987 [hep-th]].
  

\bibitem{Sugishita:2013jca} 
  S.~Sugishita and S.~Terashima,
  ``Exact Results in Supersymmetric Field Theories on Manifolds with Boundaries,''
  JHEP {\bf 1311}, 021 (2013)
  [arXiv:1308.1973 [hep-th]].

\bibitem{Yoshida:2014ssa} 
  Y.~Yoshida and K.~Sugiyama,
  ``Localization of 3d $\mathcal{N}=2$ Supersymmetric Theories on $S^1 \times D^2$,''
  arXiv:1409.6713 [hep-th].



\bibitem{FLM}
{I. B. Frenkel, J. Lepowsky and A. Meurman, 
``A Natural Representation of the Fischer-Griess Monster With the Modular Function $J$ As Character,'' 
Proc. Natl. Acad. Sci. USA {\bf 81} (1984) 3256-3260.}


\bibitem{Banados:2012ue} 
  M.~Banados, R.~Canto and S.~Theisen,
  ``The Action for higher spin black holes in three dimensions,''
  JHEP {\bf 1207}, 147 (2012)
  [arXiv:1204.5105 [hep-th]].


\bibitem{rademacher:1939}
H.~Rademacher,
``The Fourier Coefficients and the Functional Equation of the Absolute
  Modular Invariant $j(\tau)$,''
   Am.\ J.\ Math.\  {\bf 61}, 237 (1939).


\bibitem{Duncan:2009sq} 
  J.~F.~Duncan and I.~B.~Frenkel,
  ``Rademacher sums, Moonshine and Gravity,''
  Commun.\ Num.\ Theor.\ Phys.\  {\bf 5}, 849 (2011)
  [arXiv:0907.4529 [math.RT]].


\bibitem{Bouwknegt:1992wg} 
  P.~Bouwknegt and K.~Schoutens,
  ``W symmetry in conformal field theory,''
  Phys.\ Rept.\  {\bf 223}, 183 (1993)
  [hep-th/9210010].

\bibitem{Gaberdiel:2010ar}
M.~R. Gaberdiel, R.~Gopakumar, and A.~Saha, 
``Quantum $W$-symmetry in $AdS_3$," JHEP {\bf 1102} (2011) 004,
  arXiv:1009.6087.

  

  
\end{thebibliography}
\end{document}